\documentclass[twocolumn,superscriptaddress,aps,prl,floatfix,showpacs]{revtex4}
\usepackage{array}
\usepackage{amssymb}
\usepackage{amsmath}
\usepackage{graphicx}
\usepackage{multirow}
\usepackage{color}

\begin{document}

\title{Nuclear quantum effects in solids using a colored-noise thermostat}
\author{Michele Ceriotti}
\email{michele.ceriotti@phys.chem.ethz.ch}
\affiliation{Computational Science, Department of Chemistry and Applied Biosciences,
ETH Z\"urich, USI Campus, Via Giuseppe Buffi 13, CH-6900 Lugano, Switzerland}
\author{Giovanni Bussi}
\affiliation{Universit\`a di Modena e Reggio Emilia and INFM-CNR-S3, via Campi 213/A, 41100 Modena, Italy}
\author{Michele Parrinello}
\affiliation{Computational Science, Department of Chemistry and Applied Biosciences,
ETH Z\"urich, USI Campus, Via Giuseppe Buffi 13, CH-6900 Lugano, Switzerland}
\date{\today}

\begin{abstract}
We present a method, based on a non-Markovian Langevin equation, to include 
quantum corrections to the classical dynamics of ions in a quasi-harmonic system.
By properly fitting the correlation function of the noise, one can vary
the fluctuations in positions and momenta as a function of the 
vibrational frequency, and fit them so as to reproduce the quantum-mechanical behavior,
with minimal \emph{a priori} knowledge of the details of the system.
We discuss the application of the thermostat to diamond and to ice Ih.
We find that results in agreement with path-integral molecular dynamics can be obtained
using only a fraction of the computational effort.
\end{abstract}

\maketitle

Nuclear quantum effects are extremely important in many condensed-phase 
systems. For instance, zero-point fluctuations affect static correlations,
and energy quantization causes deviations from the classical value
of the specific heat at low temperatures.
A quantum treatment of the ionic degrees of freedom is 
mandatory to capture such effects, which are particularly 
important when light atoms or stiff vibrational modes are present.
However, including quantum effects is computationally demanding, 
even if one is interested only in equilibrium properties, as is the case here.
For this reason,
the nuclei in molecular simulations are often treated classically,
even when the electronic degrees of freedom are accounted for
quantum-mechanically~\cite{marx-hutt00proc}.

When exchange-symmetry effects are not relevant, the method of choice
for studying equilibrium expectation values is 
path-integrals molecular dynamics (PIMD)~\cite{cepe95rmp,marx+99nat}. 
However, this comes at a high computational cost, since many replicas
of the system need to be simulated in parallel.
Approximate but less expensive methods 
such as Feynman-Hibbs effective potentials~\cite{feynman-hibbs} have also been used,
as well as semiclassical approaches to treat zero-point energy (ZPE)~\cite{semiclassic}.
Their range of validity is limited to weak quantum behavior, and to
cases where the Hessian of the potential is available and
cheap to compute. The interest in methods to introduce 
quantum effects in classical trajectories is thus still very high, see 
e.g. Ref.~\cite{recent} for recent applications. 

In a recent work~\cite{ceri+09prl} we have demonstrated that a generalized, linear Langevin
equation with colored noise can be used to obtain a highly tunable
thermostat for constant-temperature molecular dynamics. In this Letter
we will show that, by suitably extending this idea, it is 
possible to modify Hamilton's equations so as to introduce
the quantum-mechanical effects. Furthermore, this comes with only a negligible 
increase in computational effort with respect to traditional classical simulations, 
and requires a minimal prior knowledge of the properties of the system. 
The idea can be traced back to the semiclassical approximation to the 
quantum Langevin equation~\cite{gle}, which has been used to model
quantum systems in contact with a quantum harmonic bath. 
A similar idea has been recently used by Buyukdagli et. al~\cite{buyu+08pre},
in order to compute quantum specific heat in harmonic systems.
However, Buyukdagli's scheme is only qualitatively correct even for a harmonic oscillator,
and neglects ZPE completely.
In this Letter we propose a more general approach, which allows
one to obtain high accuracy in reproducing quantum specific heats, and 
also tackles the more challenging task of introducing 
zero-point motion effects. This is achieved by effectively and automatically 
enforcing the quantum position and momentum distributions.
Our method gives excellent results in systems with limited 
anharmonic coupling, and is ideal for treating quantum effects
in solids.

Let us first consider a harmonic oscillator which is evolved in time
according to a generalized Langevin equation.
This equation can be written in a Markovian form by suitably extending the
state vector~\cite{marc-grig83jcp,gard03book,ceri+09prl}.
For each degree of freedom, a set of $n$ additional momenta $s_i$ are introduced,
which complement the position $q$ and the 
physical momentum $p$. The value of $n$ depends on the structure of the
memory kernel for the generalized Langevin equation and, in our experience, a choice between
$4$ and $12$ allows sufficient flexibility.
For simplicity, we assume mass-scaled coordinates, $q\leftarrow q\sqrt{m} $ and $p\leftarrow p/\sqrt{m}$.
We introduce a compact convention to represent 
matrices acting on the state vector $\left(q,p,\mathbf{s}\right)^T$:
\newcommand\arS{\rule{0pt}{12pt}}
\begin{equation}
\begin{array}{ccccc}
      &   q   &    p   &   \mathbf{s}  & \arS \\ \cline{2-4}
\multicolumn{1}{c|}{q} & a_{qq} & a_{qp} & \multicolumn{1}{c|}{\mathbf{a}_q^T} & \arS \\\cline{3-4}
\multicolumn{1}{c|}{p} & \multicolumn{1}{c|}{\bar{a}_{qp}} &  a_{pp} &  \multicolumn{1}{c|}{\mathbf{a}_p^T} & \arS \\\cline{4-4}
\multicolumn{1}{c|}{\mathbf{s}} &  \multicolumn{1}{c|}{\bar{\mathbf{a}}_q}  &  \multicolumn{1}{c|}{\bar{\mathbf{a}}_p} &  \multicolumn{1}{c|}{\mathbf{A}} & \arS \\\cline{2-4}
\end{array}
\hspace{-8pt}\begin{array}{cc}
\arS \\ \arS \\
\left.\rule{0pt}{12pt}\right\}\!\mathbf{A}_p \\
\end{array}
\hspace{-8pt}\begin{array}{cc}
\arS \\
\left.\rule{0pt}{20pt}\right\}\!\mathbf{A}_{qp} 
\end{array}
\label{eq:notation}
\end{equation}
Thus, a matrix without subscript acts on the subspace of additional momenta only.  The 
$p$ and $qp$ subscripts denote matrices which also act on the $p$ and on the $(q,p)$ respectively.
The Markovian form for the generalized Langevin equation reads
\begin{equation}
\left(\begin{array}{c}\dot{q} \\\dot{p}\\ \dot{\mathbf{s}}\end{array}\right)=
-\left(
\begin{array}{ccc}
0 & -1 & \mathbf{0} \\
\omega^2 & a_{pp} & \mathbf{a}_p^T \\ 
\mathbf{0} & \bar{\mathbf{a}}_p & \mathbf{A}
\end{array}\right)
\left(\begin{array}{c}q\\ p\\ \mathbf{s}\end{array}\right)+
\left(\begin{array}{ccc}
0 & 0& \mathbf{0}\\
0 & \multicolumn{2}{c}{\multirow{2}{*}{$\mathbf{B}_p$}}\\
\mathbf{0} & & \\
\end{array}\right)
\left(\begin{array}{c}0\\\multirow{2}{*}{$\boldsymbol{\xi}$} \\ \\\end{array}\right)
\label{eq:gle}
\end{equation}
where $\boldsymbol{\xi}$ is a vector of $n+1$ uncorrelated gaussian
random numbers.
Equation~(\ref{eq:gle}) has been chosen as the most general linear stochastic equation
where the position $q$ is not coupled with the noise nor with the $s_i$.
This form allows for an easier generalization to the anharmonic case.
The static covariance matrix $\mathbf{C}_{qp}$ can be obtained by solving
the matrix equation~\cite{zwan+01book,gard03book}
\begin{equation}
 \mathbf{A}_{qp} \mathbf{C}_{qp} + \mathbf{C}_{qp} \mathbf{A}_{qp}^T=\mathbf{B}_{qp}\mathbf{B}_{qp}^T
\label{eq:ext-cov}
\end{equation}

In Ref.~\cite{ceri+09prl} we have chosen $\mathbf{B}_p$ so as to obtain
 $\mathbf{C}_{p}=k_B T$ ($c_{qq}= k_B T/\omega^2$, in our units), 
which corresponds to enforcing detailed balance.
In a quantum oscillator at finite temperature the
distribution of position and momentum is still Gaussian, but its variance has a nontrivial
dependence on $\omega$,
i.e.~$\left<p^2\right>=\omega^2\left<q^2\right>=\frac{\hbar \omega}{2}\coth \frac{\hbar \omega}{2 k_B T}$.
We can therefore perform a fitting procedure, tuning $\mathbf{B}_p$ and $\mathbf{A}_p$ 
so that the $\omega$-dependence of $c_{qq}=\left<q^2\right>/\omega^2$ and 
$c_{pp}=\left<p^2\right>$ reproduces closely the exact quantum
fluctuations. The fitting procedure is not trivial, and will be discussed elsewhere.
Once a set of parameters has been found which guarantees a good fit over a
certain frequency range $\left(\omega_{\text{min}},\omega_{\text{max}}\right)$, the thermostat
automatically enforces the correct quantum fluctuations for any system whose typical vibrations 
fall within this range. Tipically, an error below a few percents can be obtained over a frequency 
range spanning several orders of magnitude.
In the harmonic limit, the momentum and position distributions 
computed using our thermostat sample the quantum distributions within the accuracy of the fit.
Furthermore, it is easy to show that the averages of any 
local operator which depends separately on positions or momenta are
also correctly computed.

In the general, anharmonic case, the coupling between position and momentum 
is nonlinear. Therefore, one needs to use a finite time step; in particular, one
proceeds by first integrating the $\left(p,\mathbf{s}\right)$ variables, 
which can be evolved using the exact finite-time propagator built from 
$\mathbf{A}_p$ and $\mathbf{B}_p$,
and then updating $q$ and $p$ by integrating Hamilton's equations~\cite{buss-parr07pre}.

\begin{figure}[btp]
\centering\includegraphics[clip,width=1.0\columnwidth]{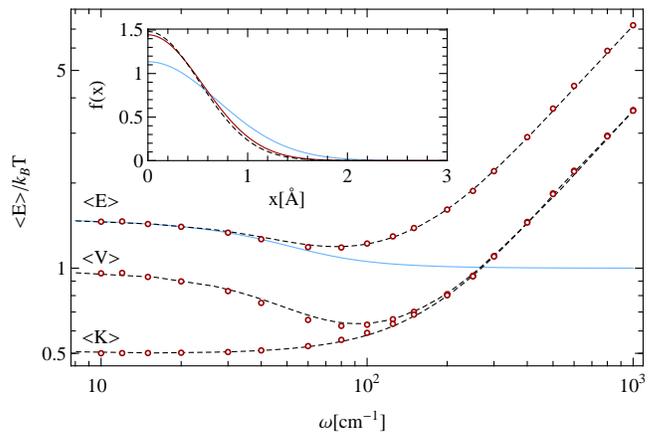} 
\caption{\label{fig:oned} (color online)
Mean total ($\left<E\right>$), potential ($\left<V\right>$)  and kinetic ($\left<K\right>$) energy
for a proton in the external potential of Eq.~\eqref{eq:oned-pot} as a function of $\omega$, for $k=1$~\AA$^{-1}$ and $T=100$~K.
In the inset, the Fourier transform of the momentum distribution is reported for the 
fully quantum, classical and colored-noise thermostat simulations at $\omega=200$~cm$^{-1}$.
Dashed black lines correspond to the exact quantum solution, 
the dark (red) series are results from present work,
using a set of parameters fitted over a frequency range from $2$~to $2000$~cm$^{-1}$.
Light (blue) lines correspond to the classical expectation values.
}
\end{figure}

As a first example, we apply this method to a one-dimensional, anharmonic potential of the form 
\begin{equation}
V\left(x\right)=\frac{\omega^2}{2m}x^2 \frac{1-e^{-kx}}{kx}
\label{eq:oned-pot}
\end{equation}
For a fixed value of $k$ and temperature $T$, this potential allows one to explore
a range of behaviors that goes from the highly anharmonic, classical limit at $\omega\rightarrow 0$
to a quantum, harmonic regime at high $\omega$. In Figure~\ref{fig:oned} we compare
the exact, quantum solution with the averages obtained using our colored-noise
thermostat. There is a remarkably good quantitative agreement not only in the asymptotic
$\omega\rightarrow 0$ and $\omega\rightarrow \infty$ limits, but also in the intermediate
region, where quantum-mechanical and anharmonic effects are significant, suggesting that both 
can be captured, albeit not fully. Also the momentum distribution (shown in the inset)
is in good agreement with its quantum-mechanical counterpart. This is particularly
appealing, since conventional PIMD can only sample positions, and one must introduce special
procedures to sample momenta as well~\cite{cepe-poll86prl,morr+07jcp,morr-car08prl}.

We now move on to more complex and realistic applications.
As we have already discussed~\cite{ceri+09prl}, if one applies a separate thermostat to each degree of freedom,
the multidimensional harmonic problem can still be solved analytically, by projecting
the dynamics onto the normal modes.  One would then expect independent phonons to 
thermalize at different effective classical temperatures according to the target $c_{qq}\left(\omega\right)$
and $c_{pp}\left(\omega\right)$ relations, provided that the response to the thermostat
is faster than the anharmonic coupling between different phonons.
Heuristically, one would expect that the error on the energy of a phonon of frequency $\omega_i$
would grow with the coupling time $\tau_H(\omega_i)$, 
as defined in Ref.~\cite{ceri+09prl},
and with the largest difference in target energy, 
$\Delta E=c_{pp}(\omega_{\text{max}})-c_{pp}(\omega_{\text{min}})$.
Any such error should instead decrease with the classical lifetime of the phonon $\tau_L(\omega_i)$,
which gauges the internal energy transfer to other vibrational modes.
This effect can be reduced to a great extent by modifying  the fit 
such that not only are the quantum distributions reproduced, but also the correlation time $\tau_H$ 
is made as small as possible.

\begin{figure}[btp]
\centering\includegraphics[clip,width=1.0\columnwidth]{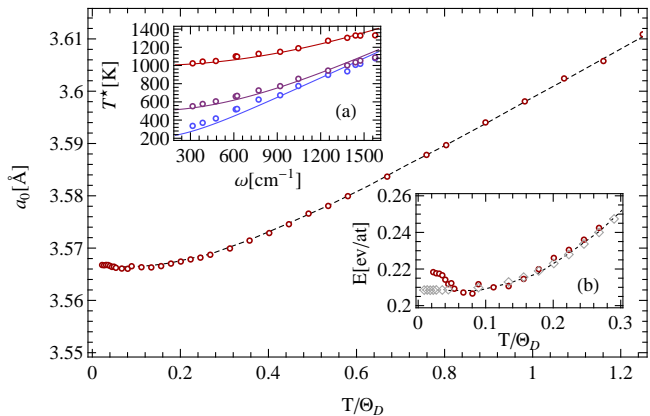} 
\caption{\label{fig:diamond} (color online)
Lattice parameter of diamond as a function of temperature, computed by isothermal-isobaric
runs at atmospheric pressure~\cite{hoov85pra}.
Black dashed lines correspond to PIMD results from
Ref.~\cite{herr-rami00prb}, and red circles to averages computed from runs using our properly
fitted colored-noise thermostat.
In inset (a) the average kinetic temperature $T^\star$, projected on a few selected normal modes,
is plotted for three target temperatures (from top to bottom,
$1000$~K, $500$~K and $200$~K) as a function of the mode frequency.
The continuous lines are the expected quantum $T^{\star}\left(\omega\right)$ curves.
In the inset (b) we compare the total internal energy (kinetic plus potential)
as computed with PIMD (black dashed line, Ref.~\cite{herr-rami00prb}) and with colored-noise thermostat fitted
to the quantum $E\left(\omega\right)$, with (red circles) and without (gray lozenges)
ZPE. The latter (gray) points have been aligned to the PIMD results at $T\rightarrow 0$. 
}
\end{figure}

 In order to assess the accuracy of this approach, we first study diamond, an archetypical 
quasi-harmonic system. We used the Tersoff classical potential~\cite{DLPOLY,ters89prb}, for which 
accurate PIMD results have been reported~\cite{herr-rami00prb}. 
In Figure~\ref{fig:diamond} we compare observables computed
with PIMD and with our colored-noise thermostat. 
The correlated Langevin dynamics is able to reproduce 
quantitatively the quantum effects on the thermal expansion
and on specific heat down to $T\approx 0.1 \Theta_D$. 
Only at lower temperatures do we start observing significant discrepancies.

To understand the reason for this breakdown, it is instructive to look at the kinetic temperature $T^\star$ of the 
different phonons [inset (a)]. This is a powerful probe of the accuracy of the
method, which can be performed whenever a normal-modes analysis is meaningful.
 In this case, the normal-modes analysis shows that, 
in the extremely quantum regime for $T<0.1\Theta_D$, the thermostat fails 
to counterbalance the phonon-phonon coupling due to anharmonicities. 
Because of this internal coupling, energy flows from the stiff modes
(which thermalize at a lower temperature than expected) to the slow ones (which 
turn out to be hotter than desired). 
A large energy difference $\Delta E\approx \hbar\left(\omega_{\text{max}}-\omega_{\text{min}}\right)/2$ 
must be maintained between fast and slow modes. Moreover, because of ZPE,
the motion of the ions is not limited to the harmonic region of the potential 
energy surface.  This leads to higher anharmonic coupling and eventually
to shorter phonon lifetimes.
This explanation is supported by the results shown in inset (b). 
Here we performed the fit by considering an energy-versus-frequency relation which
excludes the zero-point contribution, as done in Ref.~\cite{buyu+08pre}.
The agreement with PIMD energies is now virtually perfect, since 
high-frequency modes are basically frozen and normal modes are 
almost perfectly decoupled, so that internal energy transfer becomes negligible.
However, if one is interested in quasi-harmonic effects, or simply in observables
which are functions of the atomic coordinates or velocities,
neglecting ZPE would mean sampling an unphysically cold system, where stiff modes are completely 
frozen. Some examples of such observables are the Debye-Waller factor, radial distribution functions
and momentum distributions.

\begin{figure}[btp]
\centering\includegraphics[clip,width=1.0\columnwidth]{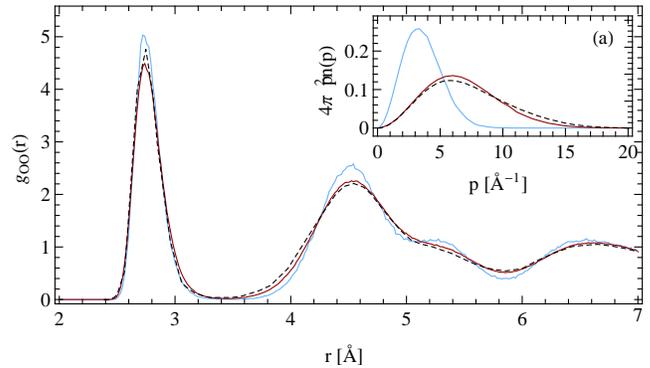} 
\caption{\label{fig:ice} (color online)
Radial distribution functions for ice at $T=220$~K. Black, dashed line corresponds
to the PIMD results of Hernandez \emph{et al.}~\cite{pena+05jcp}, dark (red) and light (blue) series to 
colored-noise and purely classical runs, respectively.
Runs have been performed in isothermal ensemble for a box of 360 water molecules
at experimental density~\cite{hayw-reim97jcp}, and are $125$~ps-long.
In inset (a), we show the experimental proton momentum distribution for ice at 
$269$~K (black dashed line)~\cite{reit+04bjp}, and corresponding distribution as computed
from a classical (blue line) and colored-noise (red line) simulation 
using a flexible TIP4P-like potential~\cite{mano-prep} at the rescaled temperature
($247$~K, i.e.~4K below the quantum melting point of the potential).
}
\end{figure}

To demonstrate the ability of our method to compute structural properties,
we have performed simulations on a TIP4P~\cite{jorg+83jcp} model of ice Ih at $220$~K,
which poses the additional challenge of showing larger anharmonicity than diamond.
In Figure~\ref{fig:ice} we show the radial distribution function $g_{OO}$
obtained using a classical simulation and our scheme,
taking as a reference literature PIMD results~\cite{pena+05jcp}.
The agreement is very good, and the peak broadening is correctly and quantitatively predicted.
Fitting the thermostat removing ZPE as in~\cite{buyu+08pre} leads to an unphysical
reduction of the width of the peaks, which are much narrower than for the classical
simulation.

We also performed similar calculations using a related flexible water forcefield~\cite{mano-prep},
in order to asses the accuracy in a system with an extreme spread of vibrational frequencies.
Results are in good qualitative agreement with rigid-water ones, but we cannot perform
a quantitative comparison in the lack of PIMD results for flexible water. 
This agreement demonstrates that a large $\Delta E$
is not a problem, provided that phonon-phonon coupling is weak.
For this model we also computed the proton-momentum distribution 
(see inset of Figure~\ref{fig:ice}). A quantitative comparison with 
experiment is difficult because of the limitations of the model potential, 
which makes it difficult to discuss the origins of the discrepancies.
For example, the simulation has to be performed  $~20$~K below the experimental temperature,
because of the dicrepancy in the melting temperature.
That said, the improvement
over the purely classical results is impressive, especially considering that it 
has been obtained at negligible computational cost.

In conclusion, we have presented a thermostatting strategy which can be applied to a vast 
class of solid-state structural problems, including disordered systems and glasses.
The approach is appealing for several reasons: no detailed information on the system is required 
(for instance, the same parameters could be used for different polymorphs of ice), and the correct
position and momentum distributions are automatically enforced. In a semiclassical sense,
it can be seen as a method to automatically equilibrate different normal modes at the 
appropriate, frequency-dependent temperature.
The implementation is straightforward, as one only needs to act on the velocities, just as with a traditional
stochastic thermostat. Most importantly, the additional computational cost is only noticeable 
for simulations employing simple, short-range, two-body potentials.
We are considering strategies to extend the range of applicability 
to extremely anharmonic systems such as liquids.
Among the possible approaches, the connections between the quantum Langevin 
equation and path-integrals formalism suggest the possibility to combine 
our method with PIMD.
Finally, we also notice that virtually any energy-versus-frequency curve
can be reproduced, so the method can be used in other applications beside 
the simulation of quantum effects.

We gratefully acknowledge David Manolopoulos for having shared with us a preprint of
his work on a flexible water potential fitted for PIMD simulations, and Gareth Tribello
for carefully reading the manuscript.

\end{document}